\documentclass[a4paper]{jpconf}
\usepackage{graphicx}
\usepackage{amsmath,amssymb}
\usepackage[LGR, T1]{fontenc}
\usepackage{babel}

\usepackage{braket}
\usepackage{cite}
\usepackage{orcidlink}
\usepackage{mathrsfs}

\def\Ob{\mathcal{O}}

\newcommand{\I}{\mathrm{i}}
\newcommand{\D}{\mathrm{d}}
\newcommand{\eq}[1]{\begin{equation}\begin{aligned}#1\end{aligned}\end{equation}}
\def\del#1#2{\frac{\partial #1}{\partial #2}}

\def\eref#1{(\ref{eq:#1})}
\renewcommand{\k}{\mathbf{k}}

\newcommand{\E}{\mathcal{E}}
\allowdisplaybreaks

\usepackage{hyperref}

\begin{document}
\title{Origin of time and probability in quantum cosmology}

\author{Leonardo Chataignier$^{1,2}\orcidlink{0000-0001-6691-3695}$,
  Claus Kiefer$^{3}\orcidlink{0000-0001-6163-9519}$, and Mritunjay
  Tyagi$^{4}\orcidlink{0009-0002-6235-8297}$} 

\address{$^1$Department of Physics and EHU Quantum Center, University of the Basque Country UPV/EHU, Barrio Sarriena s/n, 48940 Leioa, Spain}
\address{$^2$Centro Brasileiro de Pesquisas F\'{i}sicas, Rua Dr. Xavier Sigaud 150,
Urca, CEP: 22290-180, Rio de Janeiro, RJ, Brazil}
\address{$^3$Faculty of Mathematics and Natural Sciences, Institute
  for Theoretical Physics, University of Cologne, Cologne, Germany} 
\address{$^4$University of Groningen, University College Groningen,
  Hoendiepskade 23/24, 9718 BG Groningen, The Netherlands} 

\ead{leonardo.chataignier@ehu.eus, kiefer@thp.uni-koeln.de, m.tyagi@rug.nl}

\vskip 2mm

\begin{abstract}
We discuss how the classical notions of time and causal structure may
emerge together with quantum-mechanical probabilities from a
universal quantum state. For this, the process of decoherence between
semiclassical branches is important. Our discussion is based on quantum
geometrodynamics, a canonical approach to quantum
gravity. In this
framework, a particular boundary 
condition may illuminate the issue
  of the arrow of (classical) 
time in connection to the growth of entanglement entropy. 
\end{abstract}

\vspace{-0.7cm}
\section{Introduction}

If gravitation is to be described in terms of a quantum theory, a
deeper understanding of quantum foundations may be required
\cite{KieferBook}.  As a
quantum theory of gravitation would also be of relevance for cosmology,
such a theory leads to the central question
of quantum cosmology: is there a wave function of the
universe?\footnote{We refer to simplified models of the real Universe
  by using the lower-case ``universe.''} And what 
is its nature? One's view
on the matter may lead to different paths towards a theory of quanta
and the gravitational field, with distinct interpretations and
possibly different phenomenologies.

In standard quantum theory, the superposition principle plays a
  central role. If this is indeed universally valid, as experiments
  so far suggest, one arrives at what is also called the Everett
  interpretation, although what is meant is just the universal
  validity of unitary quantum theory. Unitarity refers here to the
  presence of an external time (quantum mechanics) or a non-dynamical
  spacetime background (quantum field theory). 
  There,
the quantum state $\ket{\Psi}$ can be understood as the fundamental
ontological element of the theory. Particles, fields, trajectories, observers,
observables, ``quantum jumps,'' and measurements are all to be derived
from $\ket{\Psi}$ and its dynamics, which is governed by the (possibly
functional) Schr\"odinger equation
\cite{Zeh,VaidCarroll}. More precisely, while $\ket{\Psi}$ may be a mathematical
construct, it represents the ``real structure of the world.'' This
structure paints the picture of a world governed by the superposition
principle,\footnote{Should we find a violation of this principle in an
  experiment, this view would be at least disfavored, and at
  worst discarded. Looking for such violations and their theoretical
  description is an active field of research \cite{Bassietal}.}  and
it captures the relations between different 
branches of the quantum state (the different ``waves'' that make up
$\ket{\Psi}$). Ultimately, it leads to the emergence of classical
worlds in a certain approximation. The underlying process is called
{\em decoherence} and is well tested experimentally
\cite{deco,Zeh-memory}. Although this view is without doubt
radical in its ontology, it is parsimonious in that it introduces no
modifications to unitary quantum mechanics, rather taking the
formalism as a direct representation of the world structure. 

In keeping with this ``radical conservatism,'' one can then consider
that the gravitational field is to be included in the structure
derived from $\ket{\Psi}$ and the Schr\"odinger equation, and that no
unnecessary modifications are to be introduced in the theory. As
classical general relativity exhibits spacetime diffeomorphism
invariance, it does not depend on an absolute, external notion of
time, and neither should the quantum theory. With this, the
Schr\"odinger equation must be time-independent in quantum cosmology, 
\eq{\label{eq:WDW}
0 = \I\hbar\del{}{\tau}\ket{\Psi} = \hat{H}\ket{\Psi} \ ,
}
where $\hat{H}$ is the Hamiltonian operator of gravity and matter
fields. This ``Wheeler--DeWitt'' (WDW) equation, which obviates the
necessity of an ``external'' time, stands in contrast to the usual
Everettian quantum theory, which is based on the time-dependent
Schr\"odinger equation. In quantum cosmology, classical time itself is
yet another concept that is to be derived from $\ket{\Psi}$ and the
WDW equation \cite{KieferBook,KP22}. For the gravitational part, the full
quantum state only depends on 
the {\em three-}, (not the {\em four-}) geometry.

A major challenge to Everettian quantum theory and to quantum
cosmology is the derivation or explanation of the Born rule. Whereas
there have been many approaches towards such a derivation
(e.g.,\cite{BornEverett}, see also\cite{BK22}),
the origin of probabilities in Everett's framework remains a
contentious point of 
concern. Furthermore, the absence of a preferred time in quantum
cosmology seems to further complicate the issue: what is the
explanation of quantum probabilities in this case?  

In this contribution to the workshop proceedings, we give a
preliminary discussion of how quantum cosmology may be a consistent
account of the world structure, from which classical time and its
arrow emerge along with quantum probabilities. Our brief discussion
focuses on minisuperspace toy models, which are the mechanical
theories that describe homogeneous universes, although our
considerations might apply to more general field theories if a
suitable regularization procedure can be adopted
\cite{TsamisWoodard}. 

\section{Hilbert space and probabilities}

We begin with a quantum state $\ket{\Psi}$, a Hamiltonian $\hat{H}$,
and Eq.~\eref{WDW}. These are our building blocks. Eq.~\eref{WDW}
endows the state $\ket{\Psi}$ with a ``world structure,'' as we will
discuss in Sec.~\ref{sec:world}. Because of the linearity of
Eq.~\eref{WDW}, any solution $\ket{\Psi}$ can be written in terms of a
superposition of different states. If some of these states are to
correspond to semiclassical worlds, where quantum interference effects
are suppressed at least for the macroscopic spacetime geometry, we
need a way of ascertaining the absence of interference between the
different terms in a superposition. For this, we may search for a
notion of orthogonality of states: two terms in a superposition do not
interfere (considerably) if they are (approximately) orthogonal. We
would also like to determine if a set of such non-interfering worlds
is complete, in the sense of exhausting mutually exclusive
(orthogonal) alternatives. 

Appropriate notions of completeness and orthogonality arise if we
equip the space of solutions to Eq.~\eref{WDW} with an inner product,
with which we define a Hilbert space.\footnote{We take this space to
  be a vector space over the complex numbers.} It is, of course,
  an open problem whether a Hilbert-space structure can be applied to
  full quantum gravity \cite{KieferBook}, but we shall propose in the
  following a well-motivated choice.
  
We define an inner product $\braket{\cdot|\cdot}$
such that $\hat{H}$ is symmetric, obeying
$\braket{\hat{H}\psi_2|\psi_1} = \braket{\psi_2|\hat{H}\psi_1}$, and
that it is moreover self-adjoint or can be extended to a self-adjoint
operator. In this case, there is a complete orthonormal system of
energy eigenstates, $\hat{H}\ket{\k} = E(\k)\ket{\k}$. These
eigenstates are labeled by the quantum numbers $\k$, which can be seen
as the eigenvalues of a complete set of mutually commuting operators,
and we have the relations $\braket{\k'|\k} = \delta(\k',\k)$ and
$\sum_{\k}\ket{\k}\!\!\bra{\k} = \hat{1}$, where we must replace the
summation by an integration with an appropriate measure over the $\k$
variables that are continuous (if there any).\footnote{In general,
  $\delta(a,b) = 0$ if $a \neq b$, and $\delta(a,a) = \delta(0,0)$,
  which is equal to $1$ if the variables are discrete [in this case,
  we write $\delta(a,b)=\delta_{a,b}$]. If the variables are
  continuous, $\delta(\cdot,\cdot)$ corresponds to the Dirac delta
  distribution.} The eigenvalue $E(\k)$ is, in general, a function of
the $\k$ labels. The projector onto the eigenspace of $\hat{H}$ with
eigenvalue $E(\k)=\E$ can be written as\footnote{We use the symbol
  $:=$ to specify an equality that defines the quantity on the
  left-hand side.} 
\eq{\label{eq:E-proj}
  \hat{P}_\E := \sum_{\k}\delta(E(\k),\E)\ket{\k}\!\!\bra{\k}
  =
  \int_{\mathscr{C}}\frac{\D\tau}{2\pi\hbar}\,
  \e^{\frac\I\hbar\tau(\hat{H}-\E\hat{1})}
\ ,   
} 
and it satisfies $\hat{P}_{\E'}\hat{P}_\E = \delta(\E',\E)\hat{P}_\E$. The integration domain $\mathscr{C}$ on the right-hand side expression depends on the possible values of $\E$, i.e., on the spectrum of $\hat{H}$. Thus, with $\ket{\Psi_\E} = \hat{P}_\E\ket{\psi}$ and $\ket{\Phi_{\E'}} = \hat{P}_{\E'}\ket{\phi}$, we find $\braket{\Phi_{\E'}|\Psi_\E} = \delta(\E',\E)\braket{\phi|\hat{P}_\E\psi}$. If $\E'=\E$ is a continuous variable, then $\delta(\E',\E) = \delta(0,0)$ diverges [and thus the projector in Eq.~\eref{E-proj} is improper, as its square is not equal to the projector itself]. Even so, one can define an induced (or regularized) inner product on the $\E$-eigenspace by removing the Dirac delta distribution (see also \cite{HT,Refined,Halliwell}),
\eq{\label{eq:inducedIP}
\braket{\Phi_\E|\Psi_\E}_{\rm induced} := \braket{\phi|\hat{P}_\E\psi} = \int_{\mathscr{C}}\frac{\D\tau}{2\pi\hbar}\,\braket{\phi|\e^{\frac\I\hbar\tau(\hat{H}-\E\hat{1})}\psi} \ .
}
This definition provides the desired notions of orthogonality and completeness, which follow from $\hat{H}$. As any solution $\ket{\Psi}$ to Eq.~\eref{WDW} is in the eigenspace with $\E = 0$, we denote $\ket{\Psi}\equiv\ket{\Psi_{\E=0}} = \hat{P}_{\E=0}\ket{\psi}$ and $(\Phi|\Psi):=\braket{\Phi_{\E=0}|\Psi_{\E=0}}_{\rm induced}$. This leads to the (regularized) 2-norm $|\!|\!\ket{\Psi}\!|\!| = \sqrt{(\Psi|\Psi)}$, and the Hilbert space of physical states is thus the space of solutions to Eq.~\eref{WDW} that can be normalized with this regularized inner product and norm, $|\!|\!\ket{\Psi}\!|\!|<\infty$.

Notice that the constraint equation~\eref{WDW} is preserved under the
transformations $\ket{\Psi}\to \lambda\hat{U}\ket{\Psi}$, $\hat{H}\to
\hat{U}\hat{H}\hat{U}^{\dagger}$, where $\lambda$ is a complex number
and $\hat{U}$ is unitary, $\hat{U}\hat{U}^{\dagger} =
\hat{U}^{\dagger}\hat{U} = \hat{1}$. Thus, the transformed state and
Hamiltonian are equally valid starting points to construct the
theory. These transformations imply the changes
$\ket{\k}\to\hat{U}\ket{\k}$ and
$\hat{P}_{\E=0}\to\hat{U}\hat{P}_{\E=0}\hat{U}^{\dagger}$, which in
turn lead to $\ket{\psi}\to\lambda\hat{U}\ket{\psi}$ (similarly for
the states $\ket{\Phi}:=\ket{\Phi_{\E=0}}:=\hat{P}_{\E=0}\ket{\phi}$
and $\ket{\phi}$). The unitary transformations leave the induced inner
product $(\cdot|\cdot)$ invariant,\footnote{The notion of the adjoint
  of an operator and of unitarity here refer to the original inner
  product $\braket{\cdot|\cdot}$ that is used to define the induced
  inner product in Eq.~\eref{inducedIP} by means of the insertion of
  the $\hat{P}_{\E}$ operator between the $\bra{\phi}$ and
  $\ket{\psi}$ states (we are interested in the case in which
  $\E=0$). A particular case of such transformations is given by a
  unitary operator that commutes with $\hat{H}$, for which we have $\ket{\Psi}
  = \hat{P}_{\E=0}\ket{\psi}\to \lambda\hat{U}\ket{\Psi} =
  \lambda\hat{U}\hat{P}_{\E=0}\ket{\psi} =
  \hat{P}_{\E=0}(\lambda\hat{U}\ket{\psi})$ [cf. Eq.~\eref{E-proj}]
  and $\hat{H}\to\hat{H}$ (and similarly for the states
  $\ket{\Phi}:=\ket{\Phi_{\E=0}}:=\hat{P}_{\E=0}\ket{\phi}$ and
  $\ket{\phi}$).} whereas the rescaling of $\ket{\Psi}$ by a complex
number alters its phase and norm. We are thus led to the view that, as
far as the interference and completeness of worlds are concerned
[which are described by means of the inner product structure derived
from $\hat{H}$, e.g., as in Eq.~\eref{inducedIP}], what matters are
not $\ket{\Psi}$ and $\hat{H}$ but their equivalence classes under
these transformations.  

The component of $\ket{\Psi}$ along a normalized physical state
$\ket{\Phi}$ can be obtained by means of the Hermitian projector
$\hat{P}_{\Phi} := \ket{\Phi}\!(\!\Phi|$, which is understood to act
via the induced inner product, $\ket{\mathcal{B}_{\Phi}} :=
\hat{P}_{\Phi}\ket{\Psi} = (\Phi|\Psi)\ket{\Phi}$.\footnote{Of course,
  one could, in principle, attempt to employ other notions of inner
  products and norms to compute the components of $\ket{\Psi}$ along
  other states, and one might even consider non-Hermitian
  projectors. Such arbitrary notions would, however, lack
  motivation. On the other hand, the idea here is that, given the
  starting points of the theory, which are $\hat{H}$ and $\ket{\Psi}$,
  the induced inner product $(\cdot|\cdot)$ follows from the
  Hamiltonian, insofar the given $\hat{H}$ is (or can be extended to)
  a self-adjoint operator in a suitable metric
  $\braket{\cdot|\cdot}$. Given this well-motivated inner product,
  $\ket{\mathcal{B}_{\Phi}}$ is simply the standard expression for the
  component of a vector along another vector with unit norm.} As a
matter of terminology, we call the component
$\ket{\mathcal{B}_{\Phi}}$ a ``branch state.'' The notion of branch
state as a component of $\ket{\Psi}$ can be generalized to
superpositions of components. Indeed, if
$\{\ket{\Phi_\alpha},\alpha\in\mathcal{I}\}$ is a basis\footnote{Here,
  we consider that the basis elements define a complete set of states,
  which are not necessarily orthogonal. If the $\alpha$ labels are
  continuous, it may be that the basis elements can only be normalized
  to Dirac delta distributions, in which case they form a complete
  orthornormal system of elements that are not in the Hilbert space
  (as they are not normalizable). We nevertheless still refer to such
  a system as a basis. For example, from Eqs.~\eref{E-proj},
  and~\eref{inducedIP}, we see that $\ket{\k_*}:=\ket{\k}_{E(\k)=0}$
  defines a complete set because we can write a solution to
  Eq.~\eref{WDW} as $\ket{\Psi} = \hat{P}_{\E=0}\ket{\psi} =
  \sum_\k\delta(\E(\k),0)\braket{\k|\psi}\ket{\k} \equiv
  \sum_{\k_*}\mathcal{N}^2(\k_*)\psi(\k_*)\ket{\k_*}$, where
  $\mathcal{N}^2(\k_*)$ is a real normalization factor that may appear
  after the elimination of $\delta(\E(\k),0)$. We can then define the
  basis vectors as $\ket{\Phi_{\k_*}}:=\mathcal{N}(\k_*)\ket{\k_*}$,
  with $\k_*$ playing the role of the indices $\alpha$ in
  $\{\ket{\Phi_\alpha},\alpha\in\mathcal{I}\}$. Notice that
  $\ket{\Phi_\alpha}$ are themselves solutions to Eq.~\eref{WDW}, and,
  as such, may also exhibit some interesting ``world structure'' or
  (emergent) dynamics (see Sec.~\ref{sec:world}).} in the physical
Hilbert space, with $\mathcal{I}$ being a set of indices, we can
define branch states as the superpositions 
\eq{\label{eq:branch}
\ket{\mathcal{B}_i} := \hat{P}_i\ket{\Psi} := \!\sum_{\alpha\in
  K_i\subset\mathcal{I}}\!\hat{P}_\alpha\ket{\Psi} :=
\!\sum_{\alpha\in
  K_i\subset\mathcal{I}}(\Phi_\alpha|\Psi)\ket{\Phi_\alpha} \ . 
}
If $\mathcal{I}$ is the union of all $K_i$ subsets, then the sum of
branches equals the total state,
$\sum_i\ket{\mathcal{B}_i}=\ket{\Psi}$, because of the completeness of
the $\ket{\Phi_\alpha}$ states.\footnote{The projectors
  $\hat{P}_\alpha$ along the $\ket{\Phi_\alpha}$ states may be
  improper if the basis is not orthonormal with respect to the induced
  inner product or if the $\alpha$ labels are continuous. Furthermore,
  if $\alpha$ is a continuous label, then branches can be obtained by
  integrating over an open region $K_i$ of the $\alpha$ parameter
  space with an appropriate integration measure,
  $\ket{\mathcal{B}_i}:=\int_{K_i}\D\mu_i(\alpha)\,(\Phi_\alpha|\Psi)\ket{\Phi_\alpha}$. If
  the definition of the measures $\D\mu_i(\alpha)$ tacitly includes a
  partition of unity subordinate to the open regions, and if the
  parameter space of the $\alpha$ labels can be covered by such open
  regions, then the sum of such branches gives the total state.} Just
as $\ket{\Psi}$, $\ket{\mathcal{B}_i}$ solves Eq.~\eref{WDW}, and it
is only physically meaningful up to unitary transformations and
complex rescalings. Indeed, the transformation
$\ket{\Psi}\to\lambda\hat{U}\ket{\Psi}$ is equivalent to transforming
the basis states as $\ket{\Phi_\alpha}\to\hat{U}\ket{\Phi_\alpha}$ and
the coefficients as
$(\Phi_\alpha|\Psi)\to\lambda(\Phi_\alpha|\Psi)$. With this, the
branches transform in the same way as the universal state
$\ket{\Psi}$; i.e.,
$\ket{\mathcal{B}_i}\to\lambda\hat{U}\ket{\mathcal{B}_i}$. 

Although general branches may fail to be orthogonal, the branches that correspond to semiclassical worlds must not noticeably interfere, and if they are to be states in the physical Hilbert space, which are on a par with the universal state $\ket{\Psi}$, then they must be normalizable. Thus, semiclassical worlds are to be represented by branches that satisfy
\eq{\label{eq:deco-branch}
(\mathcal{B}_i|\mathcal{B}_j) = (\Psi|\hat{P}_i\hat{P}_j\Psi) \approx \delta_{ij}(\Psi|\hat{P}_i\Psi) \ ,
}
at least approximately. This orthonormality condition yields a notion of \emph{decoherence}, albeit one introduced prior to any distinguished notion of time or probability. We will further discuss this below.

If the semiclassical world described by $\ket{\mathcal{B}_i}$
(assuming it is a nonvanishing branch) contains subsystems that
represent observers, the absence of interference with other worlds
implies that the results of experiments performed by the observers
will, for all practical purposes, be described solely by the state
$\ket{\mathcal{B}_i}$ rather than the universal state $\ket{\Psi}$. In
this way, $\ket{\mathcal{B}_i}$ takes on the role of universal state
for these observers, serving as a standalone solution to
Eq.~\eref{WDW} (this corresponds to a ``state update'' or ``apparent
collapse'' $\ket{\Psi}\to\ket{\mathcal{B}_i}$). 

The freedom to rescale solutions to Eq.~\eref{WDW} by a complex number
means that physical states are, in fact, rays in the physical Hilbert
space. (This conclusion can only be drawn for the full state of
  the whole quantum universe, which is the only closed system in the
  strict sense.)
Representatives of physical states can be fixed by a
normalization condition, such as $|\!|\!\ket{\Psi}\!|\!| = 1$, which
we adopt from now on. Thus, the universal state is normalized, and
only the unitary invariance remains. With this convention, the
observers for whom the nonvanishing branch $\ket{\mathcal{B}_i}$ is
the ``apparent'' universal state can accordingly normalize it by
defining
$\ket{\zeta_i}=\ket{\mathcal{B}_i}/|\!|\!\ket{\mathcal{B}_i}\!|\!|$ in
the regime in which Eq.~\eref{deco-branch} holds. As a matter of
notation, one can also define $\ket{\zeta_i} = 0$ if
$\ket{\mathcal{B}_i}=0$, although this trivial branch cannot have
observers, as it has no structure. 

What is the probability that observers will find themselves in the
world $\ket{\zeta_i}$? This a rather subtle and contentious issue in
the literature \cite{BornEverett}, where most works focus on the
Everettian approach to quantum mechanics based on the time-dependent
Schr\"odinger equation. In the case of quantum cosmology considered
here, which is based on the time-independent equation~\eref{WDW}, one
might expect the issue to become even more vexed. Nevertheless, one
can reason that, if we can quantify ``how much'' of $\ket{\Psi}$
corresponds to a given branch in terms of a ``percentage,'' then the
probability of (finding oneself in) a given branch will simply be this
number. 

More precisely, we search for a measure
$\mu_{\ket{\Psi}}(\ket{\zeta})$ that assigns to any normalized branch
of $\ket{\Psi}$ a real number in the interval $[0,1]$. We require that
this measure be ``non-contextual,'' i.e., that its functional form
does not depend on any particular basis of the physical Hilbert space,
as no such preferred structure enters in the definition of the theory
(the starting points are solely $\hat{H}$ and
$\ket{\Psi}$). Furthermore, $\mu_{\ket{\Psi}}(\ket{\zeta})$ should
only depend on the unitary equivalence classes,
$\mu_{\hat{U}\ket{\Psi}}(\hat{U}\ket{\zeta}) =
\mu_{\ket{\Psi}}(\ket{\zeta})$, and it is also reasonable to require
that $\mu_{\ket{\Psi}}(\ket{\zeta})$ be a continuous additive function
of the branches, $\mu_{\ket{\Psi}}(\ket{\zeta_{1+2}}) =
\mu_{\ket{\Psi}}(\ket{\zeta_1})+\mu_{\ket{\Psi}}(\ket{\zeta_2})$, with
$\ket{\zeta_{1+2}} =
\frac{\ket{\mathcal{B}_1}+\ket{\mathcal{B}_2}}{|\!|\!\ket{\mathcal{B}_1}+\ket{\mathcal{B}_2}\!|\!|}$
if at least one of the branches is nonzero or $\ket{\zeta_{1+2}}=0$ if
both vanish.\footnote{A similar requirement was adopted in the
  original works of Everett \cite{Everett}.} Intuitively, this
corresponds to the requirement that the ``portion'' of $\ket{\Psi}$
that corresponds to the branch
$\ket{\mathcal{B}_1}+\ket{\mathcal{B}_2}$ be simply obtained from the
sum of the ``portions'' associated to $\ket{\mathcal{B}_1}$ and
$\ket{\mathcal{B}_2}$, and, moreover, that a small variation in the
branch does not lead to ``jumps'' in the ``percentage''
$\mu_{\ket{\Psi}}$. In the particular case in which both branches
vanish, this additivity requirement implies that
$\mu_{\ket{\Psi}}(0)=0$ (vanishing branches are assigned zero
measure). Finally, we also require that
$\mu_{\ket{\Psi}}(\ket{\Psi})=1$ (the maximum value, $100\%$, is
reached if the branch coincides with $\ket{\Psi}$). 

To find $\mu_{\ket{\Psi}}(\ket{\zeta})$, we first note a useful result. Given $\ket{\Psi}$ and a normalized arbitrary state $\ket{\Phi}$, a continuous function $f(\ket{\Psi},\ket{\Phi})$ that is unitarily invariant [i.e., it satisfies $f(\hat{U}\ket{\Psi},\hat{U}\ket{\Phi}) = f(\ket{\Psi},\ket{\Phi})$] is of the form $f(\ket{\Psi},\ket{\Phi}) = g(z)$, where $g$ is a continuous function of the overlap $z := (\Phi|\Psi) = (\hat{U}\Phi|\hat{U}\Psi)$, which has the same (in general, complex) value for each unitary equivalence class. This can be shown following the work of Gogioso (see \cite{Gogioso} and references therein). If we consider the (sub)group $\mathcal{U}_\Psi$ of unitary transformations that leave $\ket{\Psi}$ invariant, $\hat{U}_{\Psi}\ket{\Psi} = \ket{\Psi}$, we find that $\hat{U}_\Psi \in \mathcal{U}_\Psi$ implies $\hat{U}^{\dagger}_\Psi\in\mathcal{U}_\Psi$ and vice versa. Then, $z = (\Phi|\Psi) = (\hat{U}_{\Psi}\Phi|\Psi)$ has a constant value along the orbit of $\ket{\Phi}$ under $\mathcal{U}_\Psi$, which is the set of states of the form $\hat{U}_\Psi\ket{\Phi}$ with $\hat{U}_\Psi\in\mathcal{U}_\Psi$. If another normalized physical state $\ket{\xi}$ has the overlap $(\xi|\Psi) = (\Phi|\Psi) = z$, we can write $z^*\ket{\Psi} = \ket{\Phi}-\ket{\beta} = \ket{\xi}-\ket{\gamma}$, where $\ket{\beta}$ and $\ket{\gamma}$ are orthogonal to $\ket{\Psi}$ in the induced inner product. Because of the fact that $\ket{\Psi}, \ket{\Phi}$, and $\ket{\xi}$ have unit norm, we obtain $|\!|\!\ket{\beta}\!|\!|^2 = |\!|\!\ket{\gamma}\!|\!|^2 = 1-|z|^2$. With $\rho\,\mathrm{e}^{\I\theta}:=(\beta|\gamma)$ and $\ket{\chi}:=\mathrm{e}^{\I\theta}\ket{\beta}-\ket{\gamma}$, we can construct the operator $\hat{u}:= \ket{\Psi}\!(\Psi|+\mathrm{e}^{\I\theta}\left[\hat{1}-\ket{\Psi}\!(\Psi|-\frac{2}{(\chi|\chi)}\ket{\chi}\!(\chi|\right]$. Here, $\hat{1}$ can be understood as the restriction of the identity to the physical Hilbert space. It is straightforward to see that $\hat{u}$ is a unitary operator on the physical Hilbert space that satisfies $\hat{u}\ket{\Phi} = \ket{\xi}$, and $\hat{u}\ket{\Psi} = \ket{\Psi}$, and therefore $\hat{u}\in\mathcal{U}_{\Psi}$ and $\ket{\xi}$ is in the orbit of $\ket{\Phi}$ under $\mathcal{U}_\Psi$. This means that $z$ uniquely labels an orbit, as it has a constant value over it, and two normalized states with the same value of $z$ must be on the same orbit.\footnote{Suppose that the orbits would be labeled by a set of independent parameters $(z,\eta,\ldots)$. Then, it would be possible to change orbits by holding $z$ fixed while varying the other parameters. But this would imply that different orbits could have the same value of $z$, which contradicts the above result. In this way, $z$ uniquely labels the orbits of states under $\mathcal{U}_\Psi$, without other independent parameters, given that $\ket{\Psi}, \ket{\Phi}$, and $\ket{\xi}$ have unit norm.} Finally, the unitary invariance of the function $f$ implies, in particular, that $f$ is constant along the orbit of $\ket{\Phi}$ under $\mathcal{U}_{\Psi}$, $f(\ket{\Psi},\hat{U}_{\Psi}\ket{\Phi}) = f(\ket{\Psi},\ket{\Phi})$, and thus it must be a continuous function of $z$, $f(\ket{\Psi},\ket{\Phi}) = g(z)$.

Now we can apply this result to $\ket{\Psi}$ and a normalized branch $\ket{\zeta_i}$. The unitary invariance of the measure $\mu_{\hat{U}\ket{\Psi}}(\hat{U}\ket{\zeta_i}) = \mu_{\ket{\Psi}}(\ket{\zeta_i})$ implies that it is a continuous function of $z_i := (\zeta_i|\Psi) = [(\Psi|\hat{P}_i|\Psi)]^{\frac12}$, which is a non-negative real number because $z_i^2$ is a sum of non-negative real quadratic forms [cf. Eq.~\eref{branch}], and it satisfies $0\leq z_i^2\leq 1$ because $\hat{P}_i$ is part of a complete set of projectors. With this, we can write $\mu_{\ket{\Psi}}(\ket{\zeta_i}) = g([(\Psi|\hat{P}_i\Psi)]^{\frac12})\equiv\tilde{g}((\Psi|\hat{P}_i\Psi)) = \tilde{g}(z_i^2)$, where $\tilde{g}$ is continuous and defined by composition with domain $[0,1]$. Moreover, by hypothesis, $\tilde{g}$ is bounded, as the image of the measure also coincides with the interval $[0,1]$. The requirement of additivity in the branches, which is equivalent to additivity in the projectors $\hat{P}_i$, can be fulfilled if $\tilde{g}$ is an additive function of its argument. As one can prove that bounded additive functions defined on $[0,1]$ must be linear (see, e.g., \cite{CauchyEquation}), we conclude that $\tilde{g}(z_i^2) = cz_i^2$ for some positive constant $c$. Finally, the requirement $\mu_{\ket{\Psi}}(\ket{\Psi}) = 1$ translates to $g(1) = 1$, which fixes $c=1$. We thus obtain the Born rule:
\eq{\label{eq:Born}
\mu_{\ket{\Psi}}(\ket{\zeta_i}) = (\Psi|\hat{P}_i\Psi) \ ,
}
for a universal state $\ket{\Psi}$ that is normalized in the induced inner product. This defines the ``percentage'' of the universal state that corresponds to a given branch, and we identify it with the probability that observers will find themselves in that branch (see also \cite{BornEverett}). Notice that, in general, Eq.~\eref{Born} is approximate in that it holds in the regime in which the decoherence condition in Eq.~\eref{deco-branch} is valid, because there the branches that correspond to non-interfering worlds can be normalized.

This derivation of the Born rule essentially corresponds to the
adaptation of the derivation discussed by Gogioso \cite{Gogioso}, as
well as that proposed by Everett \cite{Everett}, to the
time-reparametrization invariant setting of quantum cosmology,
governed by the WDW equation~\eref{WDW}. Many subtleties that are
discussed in the literature (e.g., in \cite{BornEverett}) may still
apply, and thus this derivation should not be seen as a definitive
proof, but rather as a strong indication that the Everettian view is
also consistent in this reparametrization invariant setting. As
Gogioso remarked \cite{Gogioso}, the derivation can be related to a
special case of Gleason's theorem \cite{Gleason} (see also the theorem
proved by Busch \cite{Busch} and the discussion in
\cite{CauchyEquation}). However, the emphasis here is not in showing
that probability distributions can be written in terms of a Born-rule
expression. Rather, the proposition is that solutions to the WDW
equation~\eref{WDW}, when taken as direct representations of worlds
and as the basic ontological elements of the theory, lead to a
Born-rule distribution of semiclassical worlds. In this sense, the
origin of probability in quantum cosmology would lie in the structure
of Eq.~\eref{WDW} and its solutions. Next, we briefly discuss the
structure of semiclassical worlds in a particular class of toy models,
and how classical time and its arrow may emerge in this picture. With
time, the notion of repeatability of measurements, and indeed of
``consistent histories,'' can emerge. 

\section{\label{sec:world}Semiclassical worlds, classical time, and
  decoherence}

Let us now set $\hbar=c=1$ and consider a
Friedmann--Lema\^{i}tre--Robertson--Walker (FLRW) model, in which
Eq.~\eref{WDW} can be written as a differential equation of the form
\cite{KieferBook} 
\eq{\label{eq:WDW-FLRW}
0 = \hat{H}\Psi = \frac{\kappa}{2}\del{^2}{\alpha^2}\Psi+\frac{V(\alpha)}{\kappa}\Psi+\hat{H}_m\!\left(\alpha;q,\frac{\hbar}{\I}\del{}{q}\right)\Psi \ ,
}
where the variable $a=a_0\e^{\alpha}$ classically corresponds to the
scale factor of the universe, $\kappa:=8\pi G$ is the gravitational coupling
constant defined from Newton's constant $G$, and $\hat{H}_m$ is the
Hamiltonian for the matter (non-geometric) degrees of freedom $q$. If
$\hat{H}_m$ is (or can be extended to) a self-adjoint operator
relative to an inner product $\braket{\cdot|\cdot}_{m}$, which may
depend parametrically on $\alpha$, then the operator $\hat{H}$ is
symmetric with respect to the inner product
$\braket{\Psi_2|\Psi_1}:=\int_{-\infty}^{\infty}\D\alpha\braket{\Psi_2(\alpha)|\Psi_1(\alpha)}_m$. In
this inner product, the states $\ket{\alpha,q}$, which satisfy
$\braket{\alpha',q'|\alpha,q} = \delta(\alpha',\alpha)\delta(q',q)$,
form a complete system of eigenstates of the scale factor and matter
fields, and they can be thought of as ``point coincidences,'' i.e.,
states that define an event.\footnote{In classical general relativity,
  diffeomorphism invariance and background independence lead to the
  view that spacetime events can be defined in terms of ``point
  coincidences.'' This means that, rather than using arbitrary and
  unphysical coordinates to designate an event, one should rather
  resort to physically defining points (``here and now, there and
  then'') from values of the fields themselves (see, e.g.,
  \cite{ChataigThesis} and references therein).} There is no structure
to these events, however, because one can freely choose the $\alpha,q$
labels, and these eigenstates are not time-reparametrization
invariant, as they do not solve Eq.~\eref{WDW-FLRW}. It is precisely
the constraint equation~\eref{WDW-FLRW} that enforces a ``world
structure'' on its solutions. Let us see how this comes about. 

Solutions to Eq.~\eref{WDW-FLRW} can be found by means of a
Wentzel--Kramers--Brillouin (WKB) expansion with respect to the
$\kappa$ parameter, see \cite{KP22,Observations} and the references
  therein, $\Psi(\alpha,q) = 
\sum_{\k_*}\exp[\I S_{\k_*}(\alpha,q)/\kappa] =: \sum_{\k_*}\exp[\I
S_{\k_*}^{(0)}(\alpha)/\kappa]\tilde{\psi}_{\k_*}(\alpha;q)$, where
$S_{\k_*}(\alpha,q):=\sum_{n=0}^{\infty}\kappa^nS_{\k_*}^{(n)}(\alpha,q)$
are complex functions of $\alpha,q$ labeled by the parameters $\k_*$,
which identify a complete set of solutions to Eq.~\eref{WDW-FLRW}, so
that $\Psi(\alpha,q):=\braket{\alpha,q|\Psi}$ is, in general, a
superposition of such solutions. We can write $\k_*=(\sigma,\k_m)$,
where the $\sigma$ label is related only to the gravitational degrees
of freedom. The lowest-order term in the $\kappa$ expansion is
$\varphi_{\sigma}:= S_{\k_*}^{(0)}$, as it can be taken to depend only
on the geometric variable $\alpha$ and on the label $\sigma$ (this
follows if, for example, the kinetic term of $\hat{H}_m$ is a
positive-definite quadratic form in the classical limit
\cite{Observations,Semiclassical}). In this way, $\varphi_{\sigma}$
solves the Hamilton--Jacobi equation\footnote{This equation
  corresponds to the no-coupling limit between matter and geometry. It
  is, in principle, possible to consider the back action of matter
  onto the geometrical variables, but it typically does not affect the
  lowest order in $\kappa$ \cite{Observations,Semiclassical}. We also
  assume that $V(\alpha)$ is positive so as to allow for nontrivial
  real solutions to Eq.~\eref{HJ}, which read $\varphi_{\sigma} =
  \pm\int\D\alpha\sqrt{2V(\alpha)}$. Thus, the $\sigma$ label simply
  corresponds to the overall choice of sign for these solutions,
  $\sigma=\pm1$. In more general models, the labels of the
  gravitational sector may be more complicated and include continuous
  degrees of freedom besides the discrete sign degeneracy.} 
\eq{\label{eq:HJ}
-\frac{1}{2}\left(\del{\varphi_{\sigma}}{\alpha}\right)^2+V(\alpha) =
0 \ , 
}
and its gradient leads to the definition of a parameter $t_{\sigma}$
via $\partial/\partial t_{\sigma} :=
-N_{\sigma}(\alpha)\partial_\alpha \varphi_{\sigma}\partial_\alpha$,
where $N_{\sigma}\neq0$ is an arbitrary function (usually called the
``lapse''). This $t_{\sigma}$ is a ``time'' variable insofar $\partial
f/\partial t_{\sigma} = -N_{\sigma}\partial_\alpha
\varphi_{\sigma}\partial_\alpha f$ is exactly the classical equation
of motion of a function $f(\alpha)
=\mathtt{f}(\alpha,p_{\alpha}=\partial_\alpha \varphi_{\sigma})$ that
can be derived from the Hamilton--Jacobi equation~\eref{HJ}. It is
worth noting that $t_{\sigma}$ is an ``intrinsic time,'' in the sense
that it is defined from some of the fields in the theory instead of
being an external, independent parameter. It is also called
``WKB time'' in this context \cite{ZehTime} because it arises from the
WKB expansion of the total wave function with respect to $\kappa$. 

The wave functions $\tilde{\psi}_{\k_*}$ encode the higher orders in $\kappa$ and the dependence on the matter fields $q$. Because of the phase transformation
\eq{\label{eq:phase-transform}
\e^{-\I \varphi_{\sigma}/\kappa}\hat{H}\e^{\I \varphi_{\sigma}/\kappa} = -\frac{\I}{N_{\sigma}}\del{}{t_{\sigma}}+\hat{H}_m+\frac{\I}{2}\del{^2\varphi_{\sigma}}{\alpha^2}+\mathcal{O}(\kappa) \ ,
}
which follows from the definition of $t_{\sigma}$ and Eq.~\eref{HJ},
we see from Eq.~\eref{WDW-FLRW} that each $\tilde{\psi}_{\k_*}$ solves
a time-dependent Schr\"odinger equation in the lowest order in
$\kappa$, where the time parameter is $t_{\sigma}$ and the effective
Hamiltonian is $N_{\sigma}[\hat{H}_m+\I\,\partial^2_\alpha
\varphi_{\sigma}/2]$. This Hamiltonian depends on the classical
gravitational solution $\alpha(t_{\sigma})$ obtained from the integral
curves of $\partial/\partial t_{\sigma}$ and Eq.~\eref{HJ}, and thus
it depends on a \emph{classical} spacetime background, which arises
from the phase factor $\varphi_{\sigma}$. For this reason, each of the
$\e^{\frac{\I}{\kappa}\varphi_{\sigma}}\tilde{\psi}_{\k_*}$ factors
can be thought of as a ``semiclassical world,'' in which the quantum
matter degrees of freedom $q$ evolve relative to the classical causal
structure of the spacetime determined from $\varphi_{\sigma}$. It is
in this sense that the constraint equation~\eref{WDW-FLRW} determines
a ``world structure'' on its solutions.\footnote{Although the FLRW toy
  model has the scale factor as the sole gravitational variable, more
  general models (which classically correspond to anisotropic or
  inhomogeneous universes) will typically exhibit a host of
  gravitational variables from which a nontrivial spacetime background
  and causal structure can emerge.} 

If we define $T_{\sigma}:=\int\D{t_{\sigma}}N_{\sigma}(t_{\sigma})$,
then, from the definition of the WKB time derivative, we see that
$\mu^2:=2\pi\partial\alpha/\partial T_{\sigma}=-2\pi\partial_\alpha
\varphi_{\sigma}$. This leads to
$\partial\log\mu^2/\partial{T_{\sigma}}=-\partial^2_\alpha{\varphi_{\sigma}}$,
which in turn implies that the right-hand side of
Eq.~\eref{phase-transform} can be written as
$\mu^{-1}(-\I\partial_{T_{\sigma}}+\hat{H}_m)\mu+\Ob(\kappa)$. From
this and Eq.~\eref{phase-transform}, we see that we can write
$\e^{\frac{\I}{\kappa}\varphi_{\sigma}}\tilde{\psi}_{\k_*} =
\hat{P}_{\E=0}\e^{\frac{\I}{\kappa}\varphi_{\sigma}}\psi_{\k_*}$,\footnote{In
  this section, $\hat{H}$, $\hat{H}_m$, and $\hat{P}_{\E=0}$ stand for
  the representation of the operators in the $\ket{\alpha,q}$ basis.}
where $\psi_{\k_*}$ is an arbitrary function that need not solve the
Schr\"odinger equation. We can write this function in terms of a
superposition of eigenstates of the operator
$-\I\partial_{T_{\sigma}}+\hat{H}_m$. We obtain
$\psi_{\k_*}(T_{\sigma},q) = \mu^{-1}\!\int\!\D\E\,\e^{\I \int\D
  T_{\sigma}(\E\hat{1}-\hat{H}_m)}{\boldsymbol\psi}_{\E}(q)$ and 
\eq{\label{eq:schro-sol}
\tilde{\psi}_{\k_*} &= \e^{-\frac{\I}{\kappa}\varphi_{\sigma}}\hat{P}_{\E=0}\e^{\frac{\I}{\kappa}\varphi_{\sigma}}\psi_{\k_*} = \frac{1}{\mu}\int_{\mathscr{C}}\frac{\D\tau}{2\pi}\,\exp\left[\I\tau\left(-\I\del{}{T_{\sigma}}+\hat{H}_m\right)\right]\mu\psi_{\k_*}+\Ob(\kappa)\\
&= \frac{1}{\mu}\int\D\E\,\delta(\E,0)\e^{\I \int\D T_{\sigma}(\E\hat{1}-\hat{H}_m)}{\boldsymbol\psi}_{\E}+\Ob(\kappa)\\
&= \frac{1}{\mu}\e^{-\I \int\D T_{\sigma}\hat{H}_m}{\boldsymbol\psi}_{\E=0}+\Ob(\kappa) \ ,
}
assuming the integrals converge and can be performed in any order. Notice that the arbitrary function $\psi_{\k_*}$ (and its transform ${\boldsymbol\psi}_\E$) may contain arbitrary powers of $1/\kappa$. The function ${\boldsymbol\psi}_{\E=0}$ serves as the arbitrary initial value for the solution $\tilde{\psi}_{\k_*}$ of the Schr\"odinger equation up to order $\kappa^0$. Similarly, the induced inner product of two states $\e^{\frac{\I}{\kappa}\varphi_{\sigma'}}\tilde{\xi}_{\k'_*}$ and $\e^{\frac{\I}{\kappa}\varphi_{\sigma}}\tilde{\psi}_{\k_*}$ reads [cf. Eqs.~\eref{inducedIP} and~\eref{schro-sol}]
\eq{\label{eq:inducedIP-pre-branch}
\braket{\xi_{\k_*'}|\e^{\frac{-\I}{\kappa}\varphi_{\sigma'}}\hat{P}_{\E=0}\e^{\frac{\I}{\kappa}\varphi_{\sigma}}\psi_{\k_*}} &= \braket{\e^{\frac{\I}{\kappa}(\varphi_{\sigma'}-\varphi_{\sigma})}\xi_{\k_*'}\mu^{-1}|\e^{-\I \int\D T_{\sigma}\hat{H}_m}{\boldsymbol\psi}_{\E=0}}+\Ob(\kappa)\\
&= \int\frac{\D\alpha}{\mu^2}\,\braket{\e^{\frac{\I}{\kappa}(\varphi_{\sigma'}-\varphi_{\sigma})}\xi_{\k_*'}\mu|\e^{-\I \int\D T_{\sigma}\hat{H}_m}{\boldsymbol\psi}_{\E=0}}_m+\Ob(\kappa)\\
&= \int\frac{\D T_{\sigma}}{2\pi}\!\braket{\e^{\frac{\I}{\kappa}(\varphi_{\sigma'}-\varphi_{\sigma})}\xi_{\k_*'}\mu|\e^{-\I \int\D T_{\sigma}\hat{H}_m}{\boldsymbol\psi}_{\E=0}}_m+\Ob(\kappa)\\
&= \braket{{\boldsymbol\xi}_{\E=0}|{\boldsymbol\psi}_{\E=0}}_m+\Ob(\kappa) = \braket{\mu\tilde{\xi}_{\k'_*\k_*}|\mu\tilde{\psi}_{\k_*}}_m+\Ob(\kappa) \ , 
}
where we used the definition of $\mu$ and of the inner product
$\braket{\cdot|\cdot}$ in terms of the matter product
$\braket{\cdot|\cdot}_m$, as well as the expansion of
$\e^{\frac{\I}{\kappa}(\varphi_{\sigma'}-\varphi_{\sigma})}\xi_{\k_*'}$
in terms of the eigenstates of the operator
$-\I\partial_{T_{\sigma}}+\hat{H}_m$ with coefficients
${\boldsymbol\xi}_{\E}$. By evolving ${\boldsymbol\xi}_{\E=0}$ in the
time $T_{\sigma}$, one obtains the function
$\mu\tilde{\xi}_{\k'_*\k_*}$, which equals $\mu\tilde{\xi}_{\k'_*}$ if
$\k'_*=\k_*$. We see that the result is simply the standard
Schr\"odinger inner product for the wave functions
$\mu\tilde{\xi}_{\k'_*\k_*}$ and $\mu\tilde{\psi}_{\k_*}$, which
evolve unitarily in WKB time with the Hamiltonian $\hat{H}_m$ on the
classical spacetime background defined from $\varphi_{\sigma}$. In
this way, quantum theory on a fixed background is recovered from the
WDW equation \cite{KieferBook,Observations,Semiclassical,LapRuba}. 

As mentioned before, semiclassical worlds should not exhibit
interference between different spacetime backgrounds. This means that
we should understand these worlds more precisely as branches defined
from the $\e^{\frac{\I}{\kappa}\varphi_{\sigma}}\tilde{\psi}_{\k_*}$
terms. Branches with different $\varphi_{\sigma}$ phase factors then
ought to be approximately orthogonal in the induced inner product. Let
us write $\k_* = (\sigma,\k_m)$ with $\k_m = (s,\varepsilon)$ denoting
matter variables that are separated into ``system'' degrees of freedom
$s$ and ``environment'' degrees of freedom
$\varepsilon$.\footnote{Depending on the construction of the model,
  the ``matter environment'' may, in fact, include perturbations of
  certain gravitational degrees of freedom, such as weak gravitational
  waves \cite{Semiclassical,ContinuousMeasurement}.} We then consider
the WDW solutions
$\e^{\frac{\I}{\kappa}\varphi_{\sigma}}\tilde{\psi}_{\sigma}^{s,\varepsilon}$,
where $\mu\tilde{\psi}_{\sigma}^{s,\varepsilon}$ form a complete
system of matter states at least up to order $\kappa^0$. In order for
these states to serve as a ``basis'' for the system and environment
degrees of freedom, we assume that they are (approximately) separable
in the $s$ and $\varepsilon$ labels, and states with different
$\varepsilon$ labels are taken to be (approximately) orthogonal in the
induced inner product; i.e., we obtain the product
$\braket{\mu\tilde{\psi}_{\k'_*\k_*}|\mu\tilde{\psi}_{\k_*}}_m+\Ob(\kappa)
\approx
\delta(\varepsilon',\varepsilon)f_{\sigma',\sigma}(s',s)+\Ob(\kappa)$
[cf. Eq.~\eref{inducedIP-pre-branch}].  With this, we can define the
branches by projecting the total state $\Psi$ onto the
$\e^{\frac{\I}{\kappa}\varphi_{\sigma}}\tilde{\psi}_{\sigma}^{s,\varepsilon}$
states and summing (integrating) over the environment\footnote{One may
  also coarse grain the index $s$ by summing (integrating) over a
  subset $K_i$ of all possible values of $s$. We assume that we can
  decompose the sum (integral) $\sum_{s} = \sum_i\sum_{s\in K_i}$ for
  disjoint $K_i$ subsets.} [cf. Eq.~\eref{branch}] 
\eq{
\ket{\mathcal{B}_{\sigma;i}} := \sum_{s\in
  K_i}\sum_\varepsilon\ket{\e^{\frac{\I}{\kappa}\varphi_{\sigma}}\tilde{\psi}_{\sigma}^{s,\varepsilon}}\!(\e^{\frac{\I}{\kappa}\varphi_{\sigma}}\tilde{\psi}_{\sigma}^{s,\varepsilon}|\Psi)
\ . 
}
The overlap of two such branches reads
\eq{\label{eq:cosmo-branch-overlap}
(\mathcal{B}_{\sigma';j}|\mathcal{B}_{\sigma;i}) &= \!\sum_{s'\in K_j}\sum_{\varepsilon'}\sum_{s\in K_i}\sum_\varepsilon\braket{\psi_{\sigma'}^{s',\varepsilon'}|\e^{-\frac{\I}{\kappa}\varphi_{\sigma'}}\hat{P}_{\E=0}\e^{\frac{\I}{\kappa}\varphi_{\sigma}}\psi_{\sigma}^{s,\varepsilon}}\braket{\e^{\frac{\I}{\kappa}\varphi_{\sigma}}\psi_{\sigma}^{s,\varepsilon}|\hat{\rho}\,\e^{\frac{\I}{\kappa}\varphi_{\sigma'}}\psi_{\sigma'}^{s',\varepsilon'}}\\
&\approx \!\sum_{s'\in K_j}\sum_{s\in K_i}
f_{\sigma',\sigma}(s',s)\rho_{\rm red}(\sigma,s;\sigma',s')
+\Ob(\kappa)\,, 
}
where we used the identity
$\e^{\frac{\I}{\kappa}\varphi_{\sigma}}\tilde{\psi}^{s,\varepsilon}_{\sigma}
=
\hat{P}_{\E=0}\e^{\frac{\I}{\kappa}\varphi_{\sigma}}\psi^{s,\varepsilon}_{\sigma}$,
and we defined the density operator $\hat{\rho} =
\ket{\Psi}\!\!\bra{\Psi}$ for the total state along with the reduced
density matrix $\rho_{\rm
  red}(\sigma,s;\sigma',s'):=\sum_\varepsilon\braket{\e^{\frac{\I}{\kappa}\varphi_{\sigma}}\psi_{\sigma}^{s,\varepsilon}|\hat{\rho}\,\e^{\frac{\I}{\kappa}\varphi_{\sigma'}}\psi_{\sigma'}^{s',\varepsilon}}$. If
$\rho_{\rm red}$ is approximately diagonal in the $\sigma',\sigma$
labels, then different semiclassical spacetime backgrounds do not
interfere,\footnote{In some models, it is possible to describe this
  decoherence of the spacetime background explicitly by considering
  its ``continuous measurement'' under higher multipole variables
  \cite{ContinuousMeasurement,BK22}.} and only $f_{\sigma,\sigma}(s',s)$,
which is the overlap of the system states in the background determined
by $\varphi_{\sigma}$, remains as a prefactor in
Eq.~\eref{cosmo-branch-overlap}. If
$f_{\sigma,\sigma}(s',s)=\delta(s',s)$ (for continuous or discrete $s$
indices), then only the elements of $\rho_{\rm red}$ with $s'=s$
appear in Eq.~\eref{cosmo-branch-overlap}. If the system states are
normalizable with $f_{\sigma,\sigma}(s,s)=1$ but not orthogonal for
$s'\neq s$, there may be an approximation in which $\rho_{\rm red}$
becomes diagonal also in the $s',s$ labels. In both cases, we obtain
from Eq.~\eref{cosmo-branch-overlap} the counterpart to the general
condition in Eq.~\eref{deco-branch},
$(\mathcal{B}_{\sigma';j}|\mathcal{B}_{\sigma;i})\approx
\delta_{\sigma',\sigma}\delta_{ji}\sum_{s\in K_i}\rho_{\rm
  red}(\sigma,s;\sigma,s)$, assuming that $K_i$ and $K_j$ are
disjoint. 

In the most general case, the orthonormality relation of the branches
is thus approximately fulfilled if the reduced density matrix is
diagonal, $\rho_{\rm red}(\sigma,s;\sigma',s') \approx
\delta_{\sigma',\sigma}\delta(s',s)\rho_{\rm
  red}(\sigma,s;\sigma,s)$. This is a decoherence condition, and the
branches are distributed according to the Born rule
[cf. Eq.~\eref{Born}]. Both the approximation up to order $\kappa^0$
(weak-coupling approximation) and the decoherence between the
spacetime backgrounds (the $\sigma', \sigma$ labels) are ``timeless''
approximations that may hold in certain regions of the parameter or
configuration space. Indeed, the weak-coupling approximation is
expected to hold at large values of the scale factor, which
classically correspond to late times (large $T_{\sigma}$). Moreover,
the decoherence of the system $s',s$ labels may be achieved for
certain values of (classical) time $T_{\sigma}$ on a definite
spacetime background. This latter process is the usual decoherence
phenomenon. With the emergent classical causal structure, the wave
functions $\tilde{\psi}_{\k_*}$ evolve approximately according to the
time-dependent Schr\"odinger equation, and the usual formalism of
quantum theory follows: observers may perform repeated experiments to
probe their branch, and the repeated apparent collapse of the wave
function will lead to branches that feature a time-ordered string of
projectors, i.e., of the form
$\hat{P}_k(T_{\sigma}^n)\ldots\hat{P}_j(T_{\sigma}^2)\hat{P}_i(T_{\sigma}^1)\ket{\Psi}$,
and this leads to the notion of consistent histories (see, e.g.,
\cite{Halliwell}; the histories can also be described via path
integrals by appropriate coarse-grainings of paths that may, in fact,
preclude the need for a classical time in the description
\cite{Hartle}, similarly to our starting points). In this regime where
a classical spacetime has emerged, the analysis of subsystems and
incomplete measurements also leads to the use of positive
operator-valued measures \cite{QuantumMeasurement}. 

\vspace{-0.2cm}
\section{Outlook}
We have briefly discussed the emergence of the probabilistic nature of
quantum theory, as well as the emergence of classical time from the
wave function of the universe and its Hamiltonian constraint. There
are many possible lines of development, ranging from a detailed
analysis of decoherence in the early Universe and the definition of a
``pointer'' basis \cite{Pointer} to the connection with other
approaches \cite{BornEverett}. It is possible that this formalism also
accommodates an explanation for the observed arrow of
time\footnote{A classic text on the arrow of time is \cite{ZehBook}.} in our
semiclassical world. Indeed, the interaction terms in $\hat{H}_m$ are
typically of the form $\e^{3\alpha}V_m$, so that matter interactions
are turned off in the limit $\alpha\to-\infty$. This motivates a
choice of boundary condition to fix the total state $\ket{\Psi}$, in
which the corresponding wave function is completely separable in that
limit. The entanglement entropy of matter and gravitational degrees of
freedom would grow in the configuration-space direction of increasing
$\alpha$, and this could lead to a ``master arrow of time'' (see
\cite{Weyl,TimeAndArrow} and references therein). We hope to address these
fascinating issues in future work. 

\ack{L.C. and C.K. thank Thomas Elze for giving them the opportunity
  to present talks on this topic at the {\em Eleventh
    International Workshop DICE 2024 -- ``Quo vadis, fisica?''} in
  Castiglioncello, Italy. The work
  of L.C. was supported by the Basque Government Grant
  \mbox{IT1628-22}, and by the Grant PID2021-123226NB-I00 (funded by
  MCIN/AEI/10.13039/501100011033 and by ``ERDF A way of making
  Europe''). It was also partly funded by the IKUR 2030 Strategy of
  the Basque Government.}

\end{document}